\documentstyle[aps,epsf]{revtex}
\begin{document}

\title{On the Free Energy Monte Carlo algorithm}

\author{M.J.~Thill\\
Racah Institute of Physics\\
The Hebrew University of Jerusalem\\
91904 Jerusalem, Israel\\[2mm]
}

\maketitle

%\address{\em (\today)}

\begin{abstract}
In this paper, I investigate more closely the recently proposed 
{\em Free Energy Monte Carlo}\/ algorithm  
that is devised in particular
for calculations where conventional Monte Carlo simulations 
struggle with ergodicity problems.
The simplest version of the proposed algorithm allows for the
determination of the entropy function of statistical systems and/or performs 
entropy sampling at sufficiently large times. I also show how 
this algorithm can be used to explore the system's energy space,
in particular for minima.

%\pacs 
{\noindent PACS numbers: 05.50.+q, 11.15.Ha, 64.60.Fr, 75.50.Hk,
87.10.+e, 02.60.Pn., 02.70.Lq, 05.20.-y\\[3mm]
}

\end{abstract}

\section{Introduction}

With fast-growing computer technology, Monte Carlo (MC) simulations have
with much success been used to study various statistical systems including
neural networks, problems in biology and chemistry, lattice-gauge
theories and optimisation problems in various areas, not to mention
statistical physics, to study the properties of phase transitions and critical
phenomena.

Most MC simulations concentrate on importance sampling for
the canonical or microcanonical Gibbs ensemble, introduced by Metropolis
{\em et al}\/ \cite{MRRTT53}. The thermodynamic average, $\overline{O}$, of an
observable $O(x)$ can be estimated \cite{F63} as
\begin{equation}\label{erg}
\overline{O}=\lim\limits_{t\to\infty}
\frac{\sum\limits_{\tau=t}^{t+t_a}O(x_\tau)P^{-1}(x_\tau)\exp(-\beta H(x_\tau))}
{\sum\limits_{\tau=t}^{t+t_a}P^{-1}(x_\tau)\exp(-\beta H(x_\tau))}\,\, ,
\end{equation}
where $x_\tau$ represents a configuration at time $\tau$ of a system with
Hamiltonian $H$, $\beta=(kT)^{-1}$ is the inverse temperature (with
Boltzmann constant $k$), $t_a$ an averaging time (with $t_a\gg 1$), and $P(x)$ a
sampling probability. If $P(x)$ is chosen to be constant, very few
samples contribute significantly to the sums in the above equation, and
a very long time is required to get a reasonable estimate of $\overline{O}$.
Importance sampling comes in if one chooses $P(x)$ as the Boltzmann
weight $\exp(-\beta H(x))$. It is generally a good sampling algorithm,
but can fail to access all the parts of the phase space in available
computer time. Indeed, in many situations,
this approach or similar ones face severe ergodicity problems if there
exist many high barriers between all the possible lowest- (or nearly-lowest-)
energy configurations, as, e.g., in certain Lennard-Jones systems
(see, e.g., \cite{VC72,TV77}) and in spin glasses (see, e.g.,
\cite{BY86,MPV,FH91,R95}), to cite only two examples from statistical physics.

To overcome these, at least for a large part, it has been suggested 
that it could be more efficient to reconstruct the Gibbs
ensemble from a simulation with other ensembles (e.g., a so-called
``multicanonical MC simulation'') than to simulate it
directly (see \cite{B95} and references therein). 
One of these approaches goes under the name of the {\em entropy sampling Monte Carlo
method}\/ (ESMC). It works as follows.
Let the probability of occurence of a configuration $x$ with energy $E$
be denoted as $P(x)$, and the probability of occurence of a state with
energy $E$ as $\tilde{P}(E)$. The term ``state'' stands here for the set of all
configurations that have the same energy. They are related to each other through
\begin{equation}
\begin{array}{lcrcl}
P(x) \propto e^{-\beta E}\,\, ,\\
\tilde{P}(E) \propto N(E) e^{-\beta E} = e^{S(E)/k-\beta E}\,\, ,
\end{array}
\end{equation}
where I have introduced the entropy, $S(E)$, of the state
with energy $E$. Their number is $N(E)$.
In the Metropolis MC method \cite{MRRTT53}, the canonical distribution of
states is obtained, along with ergodicity, by a Markovian sequence in
which the transition probabilities, $\pi(x\rightarrow x')$ and
$\pi(x'\rightarrow x)$, between a pair of configurations $x$
and $x'$ are determined by the {\em detailed balance}\/ condition
\begin{equation}
\frac{\pi(x\rightarrow x')}{\pi(x'\rightarrow x)}=\frac{\exp[-\beta
E(x')]}{\exp[-\beta E(x)]}\,\, .
\end{equation}
It can be shown rigorously that, in the case of a traditional MC 
simulation, this condition ensures a simulation of the system 
with an equilibrium distribution which is just the Gibbs
distribution \cite{vK81}. The ESMC method is, however, based on the
probability distribution of states, in which the probability of occurence
of a configuration with energy $E$ is proportional to the exponential of the 
{\em negative}\/ entropy, 
\begin{equation}
\begin{array}{lcr}
P(x) \propto e^{-S(E(x))/k}\,\, ,\\
\tilde{P}(E) \propto N(E) e^{-S(E)/k}\,\, .
\end{array}
\end{equation}
In a ESMC simulation the probability of occurence of a configuration
with energy $E$ is therefore anti-proportional to the number of
configurations with that energy. In this way, the probabilities of
occurence of all states equal the same constant. An MC algorithm that does the job
is one that is based on the detailed balance condition
\begin{equation}\label{dbs}
\frac{\pi(x\rightarrow x')}{\pi(x'\rightarrow
x)}=\frac{\exp[-S(E(x'))/k]}{\exp[-S(E(x))/k]}\,\, .
\end{equation}
In all other aspects, the formalism of the ESMC procedure follows 
then the usual Metropolis procedure. It is therewith easy to show (using
the methods exhibited, e.g., in \cite{vK81})
that the ESMC algorithm simulates the system in such a way that 
all states occur with the same probability. Hence, the 
algorithm provides for a (one-dimensional) {\em random walk 
through the system's energy space}\/. 

In this spirit, Monte Carlo sampling with respect to unconventional ensembles
has received some attention (see, e.g., \cite{B95} and references
therein) in recent years. In the ``multicanonical ensemble'' approach 
\cite{our1,our1a,our3}, one samples configurations such that the exact
reconstruction of canonical expectation values becomes feasible for a
desired temperature range. Multicanonical and related sampling 
has allowed considerable gains in situations with ``supercritical'' 
slowing down, such as
\begin{description}

\item{(i)} first order phase transitions \cite{our1,our1a,our4}, 
           (for a recent review see, e.g., \cite{Janke}),

\item{(ii)} systems with conflicting constraints, such as spin glasses
           \cite{our2a,temp,BHC,Ker1} or proteins \cite{HO,HS}.

\end{description}
The reconstruction of canonical expectation values requires 
knowledge of the entropy values of the/an important part of the
energy range (see equation (\ref{erg})), but leaves innovative 
freedom concerning the optimal shape \cite{Oxford}.
Considerable practical experience exists only for algorithms
where one samples such that:

(a) The probability density is flat in a desired energy range
$P(E) = \mbox{const}$.

(b) Each configuration of fixed energy $E$ appears with the same likelihood.

\noindent It should be noted that condition (b) is non--trivial. A simple algorithm
\cite{RC} exists to achieve (a), but which gives up (b). Exact connection
to the canonical ensemble is then lost. Such algorithms are interesting
particularly for hard optimisation problems, but may be 
unsuitable for canonical statistical physics. The present paper 
focuses on achieving (a) and (b).

To achieve a flat energy distribution, the appropriate unnormalised weight
factor in equation (\ref{erg}) is $S(E)$. However, 
before simulations, the entropy function $S(E)$ is usually
not known. Otherwise we would have solved the problem in the first place. Presumably,
reluctance about simulations with an a--priori unknown weight factor
is the main reason why the earlier umbrella sampling \cite{TV77} never
became popular in statistical physics.

In the more recent papers [9--34] 
it has been suggested to overcome this loophole by simulating 
with approximate entropy values, obtained by guessing
or a short Gibbs run, and then successively simulating with ever
better estimates of the real entropy. For, if an incorrect entropy
function, $J(E)$, is used in equation (\ref{dbs}) instead of $S(E)$,
then the states with $J(E)<S(E)$ will have a larger probability to
occur than in an exact ESMC simulation, and, therefore, will be sampled
more frequently; similarly, those states with $S(E)<J(E)$ will have a
smaller one and will be sampled
less frequently. Because the entropy function, $S(E)$, is proportional
to the logarithm of the number of configurations in the corresponding
states, which in turn is proportional to the number of visits to the
states expected by the simulation, an iterative process in which  
runs with ever better $J_n(E)$ are construed can be constructed. In practice, one
simulates with $J_1(E)$ for an ``appropriately'' long time, then
$J_2(E)$ is constructed, one simulates with $J_2(E)$ for some time,
etc, and one hopes, supported by
intuition, that the $J_n(E)$ gradually approach the exact $S(E)$.
This assumption has, however, up to now never been shown 
to hold rigorously, nor is there anything known on 
convergence properties. 

Nevertheless, for first order phase transitions in non-random systems,
like ferromagnets, the problem of the a--priori unknown
weight factor in the above algorithms 
is rather elegantly overcome by means of finite size
scaling methods \cite{our1,our1a,our4,Julich,BNB,Janke}. A sufficiently
accurate estimate is obtained by extrapolation from the already simulated
smaller lattices. The smallest lattices allow still for efficient
canonical simulations. For systems with conflicting constraints the situation is less
satisfactory. Considerable attention ``by hand'' may be needed.

In this paper, I investigate more closely the recently proposed 
{\em Free Energy Monte Carlo}\/ algorithm \cite{t97l} from which the 
entropy can be obtained in the large-time limit without attention ``by hand''. 
I have applied the algorithm to the
infinite-range, the two-dimensional and the three-dimensional
ferromagnet. Where possible, I have compared the results to
the values that are known exactly. One further application of the
algorithm is the surmounting of energy barriers and/or a thorough
exploration of the system's energy space. In this paper, 
I concentrate on magnetic systems,
which have a discrete energy spectrum. However, the algorithm is
easily extended to other systems (whose energy spectrum would have
to be discretised for the use of a computer anyway).

In the next section, I describe the algorithm and show that, under
certain conditions, the algorithm converges indeed towards the correct
entropy distribution. Numerical results are exhibited in the third
section. I conclude finally in the last section with some outlooks
and further comments concerning the application of this algorithm 
to systems other than magnets.

\section{The Free Energy Monte Carlo algorithm}

\subsection{The algorithm}

I will now specify the algorithm which achieves the goal of obtaining 
the entropy function in the large-time limit. For definiteness, I will
formulate the algorithm for Ising spin systems, containing a total
number of $N$ spins. 

I will enumerate the different states of the system by their energy,
going from smallest to largest in value. Let $E(m)$ denote the
energy of the state with label $m$, say $m=0,1,\ldots,{\cal N}-1$ (so that
$\cal N$ is the total number of energy levels). Let furthermore $S(m;t)$ be the 
(estimated) entropy of the state with label $m$, at time $t$. 
At time $t=0$, I initialise $S(m;0)=0$ for all $m$. We will see below
that only entropy differences matter in the algorithm so that the
initialisation constant can in principle be chosen arbitrarily. The 
initialisation to zero is, however, preferable in the actual
implementation of the algorithm on a computer, as mentioned below.

The {\em Free Energy Monte Carlo}\/ (FEMC) algorithm then works as follows. 
Let us assume that, at time $t$, the system is in configuration $x$,
with energy $E(x)$, i.e., with label $m(x)$. 
Then, go through the following steps at time $t+1$:

\begin{enumerate}

\item Select one spin index $i$ for which the spin $s_i$ is considered
for flipping ($s_i\rightarrow -s_i$).

\item Calculate the transition probability 
\begin{equation}\label{tp}
\pi(x\rightarrow x';t+1):=\frac{1}{2}\left(1-\mbox{tanh}
\frac{S(m(x);t)-S(m(x');t)}{2}\right)
\end{equation}
to pass from configuration $x$ to configuration $x'$ which is obtained from $x$ 
by effectuating the considered spin flip. [In this present form the
algorithm is still rather an ``entropy Monte Carlo'' than a ``free
energy Monte Carlo'' algorithm; I will dwell on the full version of
the latter below.]

\item Draw a random number, $r$, uniformly distributed between zero and one.

\item If $r<\pi(x\rightarrow x';t+1)$, flip the spin, otherwise do not
flip it. In any case, the configuration of spins obtained at the end of
step 4.~is counted as the ``new configuration'', $x_{\mbox{\tiny update}}$. 

\item Now update the values of the entropy ($\mu=0,1,\ldots,{\cal N}-1$):
\begin{equation}\label{step5}
S(\mu;t+1) := S(\mu;t)+\epsilon(t) \delta_{\mu,m(x)}\,\, ,
\end{equation}
where $\epsilon(t)$ is a pre-chosen positive function which is
sufficiently small in the large-time limit, $m(x)$ the label of energy
of the configuration $x$, and $\delta$ denotes the Kronecker symbol. 

\item Go to 1. or end.

\end{enumerate}

Let us just note here that  
by the choice of $\pi(x\rightarrow x';t+1)$ as above, I ensured that
the simulation verifies a detailed balance condition ``locally'', i.e.,
at every time step. Furthermore, in the proof of the convergence of the
algorithm below, I will use a different update rule for the
entropy values, namely,
\begin{equation}\label{step5a}
S(\mu;t+1)=S(\mu;t)+\epsilon(t+1) \delta_{\mu,m(x)}
-\log\left(1+\frac{e^{S(m(x);t)}(e^{\epsilon(t+1)}-1)}{2^N}\right)\,\, .
\end{equation}
In this latter case, I shall also assume that 
the sum over $m$ of $\exp(S(m;t))$ is normalised to
equal the total number of configurations, $2^N$, at time $t=0$. 
The choice of the update rule, equation (\ref{step5a}), then ensures
that this normalisation holds for all times, $t$. If $\epsilon(t)$ 
is chosen to be small, then $S(m;t)$ can take on 
quasi-continuously values between $-\infty$ and
$N\log 2$, i.e., $S(m;t)\in [-\infty,N\log 2]$ and the corresponding
(estimated) number of configurations is bounded to values between $0$ and $2^N$. 
The interchanging of the less calculation intensive update rule, equation 
(\ref{step5}), with equation (\ref{step5a}) is possible because only 
differences in entropy enter into the calculation of the transition
probability. After every time step, one should imagine,
using equation (\ref{step5}), that the ``zero'' (baseline of the entropy
values) has been shifted upwards. The entropy may then be obtained
as a time average over instantaneous entropy values, and normalising
these values with the help of the total number of configurations, $2^N$.
The only problem one has to cope with, when using the original algorithm,
is that one may have to substract the ``baseline'' from all entropy values
every so often to avoid overflows or numerical inaccuracies when
substracting large numbers of equal order of magnitude from each other.

The above version of the algorithm does not deserve to be called
an FEMC algorithm just yet, as only the entropy enters.
However, if one wants to detect the minima (or maxima)
in the energy space, it may be useful to change the transition probability 
of the algorithm to read:
\begin{equation}\label{FEMC}
\pi(x\rightarrow x';t+1):=\frac{1}{2}\left(1-\mbox{tanh}
\frac{S(m(x);t)-S(m(x');t)-\beta (E(x)-E(x'))}{2}\right)\,\, ,
\end{equation}
where now the transition probability does not depend solely on
(instantaneous) entropy differences, but on (instantaneous) {\em free
energy}\/ differences, $\beta$ being the ``inverse temperature'' as usual.
If $\beta$ is large, the temperature is small, and the system stays
preferably in configurations with low energy (if one were to take $\beta$
small or even negative (!) one stays obviously preferably in
configurations with high energy). This can be illustrated
in particular at the beginning of the algorithm, when the entropy 
differences are zero, and one performs a gradient descent algorithm
towards a local or global minimum out of which one is then taken
by a gradual increase in (instantaneous) entropy differences. 
I will dwell on this aspect further below. With this replacement
of the transition probability, the Monte Carlo algorithm can now truly
be called {\em Free Energy Monte Carlo (FEMC) algorithm }\/.
In the next subsection, I will, however, consider again the $\beta=0$
case only, for simplicity.

\subsection{The master equation and convergence properties}

I will now write down a master equation for the probability, 
$P(S,m;t)$, to be at time $t$ in a state with
energy $E(m)$ and an (estimated) entropy function, $S=S(\mu)$
($\mu=0,1,\ldots,{\cal N}-1$).
By introducing a dependence on $S$, we
will avoid a time dependence in the transition probabilities in the 
master equation below. Furthermore, in this notation (and using the
version of the algorithm with equation (\ref{step5a}) where the
estimated values for the number of configurations per energy is 
bounded), the fact that 
the convergence of the master equation to a (unique) limit-distribution 
is ensured mathematically rigorously is easily noticed (using the
methods of \cite{vK81}).
In the following, I will also write $\pi(S,m\pm 1\rightarrow m)$
for the transition probability of moving from a configuration with
energy $E(m\pm 1)$ to one with energy $E(m)$, given that at that
time, the estimated entropy function is $S$. For simplicity,
I will consider the specific case of the infinite-range ferromagnet
in which there are only transitions between configurations where
the corresponding states are ``nearest-neighbours'' in energy space.
Furthermore, I will take $\epsilon$ to be small, but constant.
The generalisation to other systems is straightforward.

\subsubsection{The infinite-range ferromagnet}

I consider an infinite-range Ising ferromagnet consisting of $N$ spins,
$s_i$ ($i=1,2,\ldots,N$), connected by bonds of strength $J$ (I will
set $J=1$ in what follows for simplicity). The system's Hamiltonian,
${\cal H}_N$, therefore reads
\begin{equation}
{\cal{H}}_N = -\frac{1}{N}\sum\limits_{j=i+1}^N\sum\limits_{i=1}^{N-1} s_is_j
\end{equation}
As usual, the possible values are $s_i=\pm 1$. I will say that a spin, 
$s_i$, points upwards if $s_i=+1$ and downwards otherwise. 
The number, $m$, of upwards pointing spins is 
\begin{equation}
m=\sum\limits_{i=1}^{N}\frac{1}{2}(1+s_i)\,\, .
\end{equation}
I will enumerate the different states of the system by their energy,
going from smallest to largest in value. By symmetry, all configurations
with $m$ or $N-m$ upwards pointing spins have the same energy  
\begin{equation}\label{el}
E(m)=\frac{-N(N-1)+2m(N-m)}{2N}\,\, .
\end{equation}  
The total number of energy levels, ${\cal{N}}$, is therefore 
\begin{equation}
{\cal{N}}:= \left[\frac{N+1}{2}\right]\,\, .
\end{equation}
Here the (Gauss) brackets, $[x]$, indicate the smallest integer 
equal or larger than $x$, as usual. So, the variable $m$ in equation
(\ref{el}) runs from $0$ to ${{\cal{N}}-1}$. 
It is finally to note that, to every energy level, $E(m)$, corresponds 
an entropy, $S(m)$, that is easily calculated for 
the infinite-range Ising ferromagnet.
By symmetry, the number of configurations, $\omega_m$, with energy 
$E(m)$ is twice the Bernoulli number obtained by selecting $m$ spins out
of the $N$, if $m<{\cal{N}}$, and equals the Bernoulli number for
$m={\cal{N}}$, i.e.,
\begin{equation}\label{defomega}
\omega_m=\left\{
\begin{array}{cl}
2\left(
\begin{array}{c}
N\\
m
\end{array}
\right) \qquad{}&, \mbox{if  } m<{\cal{N}}\,\, ,\\
(1+\delta_{2{\cal N},N+1})\left(
\begin{array}{c}
N\\
{\cal N}
\end{array}
\right) \qquad{}&, \mbox{if  } m={\cal{N}}\,\, .
\end{array}
\right.
\end{equation}
The entropy values are therefrom obtained by taking the 
natural logarithm (I set $k=1$ here and from now on for simplicity).

\subsubsection{The master equation}

The probability distribution $P(S,m;t)$ (where $m=0,1,\ldots,{\cal
N}-1$) evolves in one time step $\Delta t$ (chosen here to equal $1$) 
as follows
\begin{equation}\label{tdt}
\begin{array}{lcl}
P(S,m;t+\Delta t) &=&\,\,\, \frac{N-m+1}{N}\,\pi(
S^{(m-1)},m-1\rightarrow m) P(
S^{(m-1)},m-1;t)\Delta t\\
{}&{}&+\, \frac{m+1}{N}\,\pi(
S^{(m+1)},m+1\rightarrow m) P(
S^{(m+1)},m+1;t)\Delta t\\
{}&{}&-\, \frac{m}{N}\, \pi(S,m\rightarrow m-1) P(
S,m;t)\Delta t\\
{}&{}&-\, \frac{N-m}{N}\, \pi(S,m\rightarrow m+1) P(
S,m;t)\Delta t\\
{}&{}&+\, P(S,m;t)
\end{array}
\end{equation}
The first two terms come from the probability of moving between time $t$
and $t+\Delta t$ into the state with energy $E(m)$ with estimated
entropy function $S$ and the other
ones from the probability of not leaving it. The functions $S^{(m\pm 1)}$ 
differ from $S$ by the
values obtained from applying the algorithm at time $t$ to a
configuration with label $m\pm 1$ to the function 
$S^{(m\pm 1)}$ and getting $S$, i.e.,
\begin{equation}\label{omuo}
S(\mu)=S^{(m\pm 1)}(\mu) + \epsilon
\delta_{\mu,m\pm 1}
-\log\left(1+\frac{e^{S^{(m\pm 1)}(m\pm
1)}(e^{\epsilon}-1)}{2^N}\right)\,\, ,\qquad
(\mu=0,1,\ldots,{\cal N}-1)\,\, .
\end{equation}
The factors in front of the transition probabilities reflect the number of spins
which can be flipped in the configuration at time $t$ to obtain a 
configuration with energy $E(m)$. More generally, they reflect they
ratio of the respective volume fraction of the total configuration space.
It is tacitly understood that for $m=0$ and $m={\cal N}-1$ 
the appropriate expressions in the equation above are zero.

To see whether the algorithm is really providing convergence towards the
correct entropy function, I define yet another probability
distribution, the probability $P(S;t)$ to obtain at
time $t$ an entropy function $S$. This probability
distribution should become a delta-peak on the true entropy function
 if one lets the time, $t$, go to infinity
and then takes the limit $\epsilon\to 0$. This amounts to our goal that 
the algorithm gives a good approximation of the true entropy in the 
infinite-time limit. Let therefore
\begin{equation}
P(S;t):= \sum_{m=0}^{{\cal N}-1} P(S,m;t)\,\, .
\end{equation}
Using equation (\ref{tdt}), the time evolution of this probability
distribution can be written down, 
\begin{equation}\label{evpot}
\begin{array}{lcl}
P(S;t+1) &=&\,\,\,
\sum\limits_{m=1}^{{\cal N}-1} \frac{N-m+1}{N}\,\pi(S^{(m-1)},m-1\rightarrow m) P(
S^{(m-1)},m-1;t)\\
{}&{}&+\, \sum\limits_{m=0}^{{\cal N}-2} \frac{m+1}{N}\,\pi(
S^{(m+1)},m+1\rightarrow m) P(S^{(m+1)},m+1;t)\\
{}&{}&-\,\sum\limits_{m=1}^{{\cal N}-1}\frac{m}{N}\,
\pi(S,m\rightarrow m-1) P(S,m;t)\\
{}&{}&-\,\sum\limits_{m=0}^{{\cal N}-2}\frac{N-m}{N}
\,\pi(S,m\rightarrow m+1) P(S,m;t)\\
{}&{}&+P(S;t)\qquad .
\end{array}
\end{equation}
I will now consider this equation in the infinite-time limit. Further
convergence properties, which are easily obtained analytically in the
case of the infinite-range ferromagnet, shall be published elsewhere~\cite{t97}.

\subsubsection{The convergence towards the true entropy function}

As mentioned in the previous subsection, the probability 
distribution $P(S,m;t)$ converges, for $t\to\infty$, to a stationary distribution, 
$P_{\mbox{\tiny eq}}(S,m)$. Therefore, also
$P(S;t)$ converges to a stationary distribution,
$P_{\mbox{\tiny eq}}(S)$. This distribution verifies
the equation
\begin{equation}\label{stpon}
\begin{array}{lcl}
0 &=&\,\,\,\,\,\, \sum\limits_{m=1}^{{\cal N}-1}
\frac{N-m+1}{N}\,\pi(S^{(m-1)},m-1\rightarrow m) P_{\mbox{\tiny eq}}(
S^{(m-1)},m-1)\\
{}&{}&+\, \sum\limits_{m=0}^{{\cal N}-2}\frac{m+1}{N}\, \pi(
S^{(m+1)},m+1\rightarrow m) P_{\mbox{\tiny eq}} (S^{(m+1)},m+1)\\
{}&{}&-\,\sum\limits_{m=1}^{{\cal N}-1}\frac{m}{N}\,\pi(S,m\rightarrow
m-1) P_{\mbox{\tiny eq}} (S,m)\\
{}&{}&-\,\sum\limits_{m=0}^{{\cal N}-2}\frac{N-m}{N}\,\pi(S,m\rightarrow
m+1) P_{\mbox{\tiny eq}} (S,m)\,\, ,
\end{array}
\end{equation}
and we shall see in the following 
that after developing this equation in orders of $\epsilon$ 
and then taking the $\epsilon\to 0$-limit, the estimated entropy
function coincides with the true entropy function. [Note here that
the limit $t\to\infty$ and $\epsilon\to 0$ are not interchangeable. 
Taking $\epsilon\to 0 $ first is the limit of performing a random
walk in configuration space~!]

At time $t$, we have the following update
\begin{equation}
S(m;t+1)=S(m;t)+\epsilon\delta_{m,\nu(t)}-\log\left(1+\frac{e^{S(\nu(t))}}{2^N}
(e^\epsilon-1)\right)\,\, ,\qquad m=0,1,\ldots,{\cal N}-1\,\, ,
\end{equation}
and I have written $\nu(t)$ to indicate the energy level of the 
configuration at time $t$. As $\nu(t)$, and therewith also $S(m;t)$ for
all $m$, is the result of a stochastic process, both $\nu(t)$ and
$S(m;t)$ for all $m$ are random variables. Let us therefore define 
$P(\nu;t)$ to be the probability of being in a configuration 
with energy $E(\nu)$ at time $t$, and develop it in the
infinite-time limit [note again that the limit exists~!]
in orders of $\epsilon$,
\begin{equation}\label{limp}
\lim\limits_{t\to\infty} P(\nu;t)= p^{(0)}(\nu)+\epsilon p^{(1)}(\nu)
+{\cal O}(\epsilon^2)\,\, ,\qquad \epsilon\to 0\,\, .
\end{equation} 

The change in entropy $\Delta S(m;t+1,t)$ between time $t$ and time $t+1$ is
\begin{equation}\label{ds}
\Delta S(m;t+1,t):= S(m;t+1)-S(m;t)
=\epsilon\left[\delta_{m,\nu(t)} -\frac{e^{S(\nu(t))}}{2^N}\right]+{\cal
O}(\epsilon^2)\,\, ,\qquad \epsilon\to 0\,\, .
\end{equation}
Now, we know that $\lim\limits_{t\to\infty}P(S,m;t)$ exists, so that
we have the stationarity condition
\begin{equation}\label{sc}
\lim\limits_{t\to\infty}<\Delta S(m;t+1,t)>_{t_a}= 0\,\, ,
\end{equation}
where $<\cdots>_{t_a}$ denotes a time average (over time $t_a$), with
$t_a\to\infty$, while holding $\frac{t_a}{t}\ll 1$ fixed. Indeed, this
average in the infinite-time-limit is nothing but the same as 
the result of a simulation at equilibrium.

Using equation (\ref{ds}) and developing the left-hand-side of equation
(\ref{sc}) in orders of $\epsilon$,
\begin{equation}
\lim\limits_{t\to \infty}<S(m;t)>_{t_a}=S^{(0)}(m)+\epsilon S^{(1)}(m)+{\cal
O}(\epsilon^2)\,\, ,\qquad \epsilon\to 0\,\, ,
\end{equation}
we can rewrite the stationarity condition, (\ref{sc}), as
\begin{equation}
0=\lim\limits_{t\to\infty}<\Delta S(m;t+1,t)>_{t_a}=
\epsilon\lim\limits_{t\to\infty}\left[<\delta_{m,\nu(t)}>_{t_a} 
-\frac{<e^{S(\nu(t))}>_{t_a}}{2^N}\right]+{\cal
O}(\epsilon^2)\,\, ,\qquad \epsilon\to 0\,\, ,
\end{equation}
and the right-hand-side must vanish order by order, in particular,
\begin{equation}\label{o1}
\lim\limits_{t\to\infty}<\delta_{m,\nu(t)}>_{t_a} 
=\lim\limits_{t\to\infty}\frac{<e^{S(\nu(t))}>_{t_a}}{2^N}\,\, ,
\end{equation}
where I just have to emphasise again that the limits on both sides
of equation (\ref{o1}) exist. Note now that the right-hand-side of
equation (\ref{o1}) is independent of $m$, and therefore a constant,
and that the left-hand-side of equation (\ref{o1}), 
using equation (\ref{limp}), equals
\begin{equation}
\lim\limits_{t\to\infty}<\delta_{m,\nu(t)}>_{t_a} =p^{(0)}(m)+{\cal
O}(\epsilon)\,\, ,\qquad \epsilon\to 0\,\, .
\end{equation}
By normalisation, we hence have
\begin{equation}
P(\nu):=\lim\limits_{t\to\infty}<P(\nu;t)>_{t_a} = \frac{1}{{\cal N}}
+{\cal O}(\epsilon)\,\, ,\qquad \epsilon\to 0\,\, ,\qquad
\nu=0,1,\ldots,{\cal N}-1\,\, .
\end{equation}
This means that in the infinite-time limit, all energy levels are
equally probable to occur. This fact is at the origin of the possibility
to replace equation (\ref{step5}) by equation (\ref{step5a}) in the update
of the algorithm in step 5., ensuring that indeed on average, only
the ``zero''/the baseline increases. Furthermore, if the algorithm is 
local in the sense that ``neighbouring'' configurations imply ``neighbouring''
energies, then the typical time to reach an energy level, $m_f$,
starting from another one, $m_o$, scales like $(m_f-m_o)^2$, as 
a general result from what is known about one-dimensional random walks.

We still have to show that
the equilibrium values for the (estimated) entropy function obtained
through the simulation do indeed, to zeroth order in $\epsilon$, equal
the true entropy values ot the system. Let us first note that
\begin{equation}
P(\nu)=\sum\limits_{\{s_i\}\big|E(\{s_i\})=E(\nu)}
P(\{s_i\})=\omega_\nu P(\{s_i\}\big|E(\{s_i\})=E(\nu))
\end{equation}
where $P(\{s_i\})$ is the equilibrium probability for the configuration
$\{s_i\}$, with energy $E(\{s_i\})=E(\nu)$, to occur. In the following, we
will also write the shorthand $P(\nu)$ for the probability of being in
a configuration $\{s_i\}$, with energy $E(\{s_i\})=E(\nu)$, and
$P(\nu;t)$ when considering time-dependent properties, respectively. Furthermore, 
by definition, we have
\begin{equation}
\omega_\nu=\frac{e^{S_0(\nu)}}{2^N}\,\, ,
\end{equation}
where $S_0(\nu)$ is the true entropy of level $\nu$ and therefore
\begin{equation}
P(\nu)=\frac{2^N}{{\cal N}e^{S_0(\nu)}}\,\, .
\end{equation}
The master equation (\ref{evpot}) for the considered algorithm reads in the 
considered infinite-time limit
\begin{equation}
\begin{array}{lcl}
0 &=&\,\,\,\,\,\, \lim\limits_{t\to\infty} \left(\,
<\frac{N-m+1}{N}\pi(S^{(m-1)},m-1\rightarrow m) P(m-1;t)>_{t_a}\right.\\
{}&{}&\qquad +\,<\frac{m+1}{N}\pi(S^{(m+1)},m+1\rightarrow m) P(m+1;t)>_{t_a}\\
{}&{}&\qquad -\, <\frac{m}{N}\pi(S,m\rightarrow m-1) P(m;t)>_{t_a}\\
{}&{}&\qquad -\,\left. <\frac{N-m}{N}\pi(S,m\rightarrow m+1) P(m;t)>_{t_a}\right)\,\, ,
\end{array}
\end{equation}
i.e., using equation (\ref{tp}) for the transition probabilities
and developing in orders of $\epsilon$,
\begin{equation}\label{sysso}
\begin{array}{lcl}
0 &=&\,\,\,\,\,\,\, \frac{N-m+1}{N}\frac{e^{S^{(0)}(m-1)-S^{(0)}(m)}}
{1+e^{S^{(0)}(m-1)-S^{(0)}(m)}}\, e^{-S_0(m-1)}
+\, \frac{m+1}{N}\frac{e^{S^{(0)}(m+1)-S^{(0)}(m)}}
{1+e^{S^{(0)}(m+1)-S^{(0)}(m)}}\, e^{-S_0(m+1)}\\
{}&{}&-\, \frac{m}{N}\frac{e^{S^{(0)}(m)-S^{(0)}(m-1)}}
{1+e^{S^{(0)}(m)-S^{(0)}(m-1)}}\, e^{-S_0(m)}
-\, \frac{m}{N}\frac{e^{S^{(0)}(m)-S^{(0)}(m+1)}}
{1+e^{S^{(0)}(m)-S^{(0)}(m+1)}}\, e^{-S_0(m)}\\
{}&{}&+\, {\cal O}(\epsilon)\,\, 
,\qquad \epsilon\to 0\,\, ,\qquad m=1,\ldots,{\cal N}-2\,\, ,
\end{array}
\end{equation}
and similarly the equations for $m=0$ and $m={\cal N}-1$.\\
Defining the difference $\delta S^{(0)}(m) := S^{(0)}(m-1)-S^{(0)}(m)$ for 
$m=0,\ldots,{\cal N}-2$ and using $\exp(S_0(m))=\omega_m$ in
the above system of equations, solving for $\delta S^{(0)}(0)$, then
$\delta S^{(1)}$, and so forth, it is easy to  see that the unique solution 
of the system (\ref{sysso}) is indeed 
\begin{equation}
S^{(0)}(m) =S_0(m)\,\, .
\end{equation}
It is likewise easy to see that with this identity the first and the third, and
the second and the fourth, terms in every equation of (\ref{sysso}) 
cancel each other, respectively, by detailed balance. Hence, to lowest
order in $\epsilon$, the estimated entropy function indeed coincides with the
true entropy function of the system.

\section{Numerical results}

I have applied the algorithm to the infinite-range, the
two-dimensional and the three-dimensional ferromagnet to see 
how well the simplest version of the algorithm (the ``$\beta=0$ 
FEMC algorithm'') performs on obtaining the entropy values of the
considered systems and on overcoming energy barriers.
More specifically, in the case of the infinite-range ferromagnet, where
I know the entropy (or number of configurations) exactly,
I have investigated the convergence to the correct values
of the entropy as well as the passage times (``tunneling times'') 
between the ``all-spins up'' and ``all-spins down'' ground states.
In the case of the two- and three-dimensional ferromagnets,
I have also studied these passage times in order to see
whether the algorithm really provides for a quick ``tunneling''
through the energy barriers. A more thorough study on the r\^ole
of the ratio of $\epsilon$ and $\beta$ in the truly FEMC algorithm
on the tunneling times shall be published elsewhere \cite{t97}.
Last but not least, I have 
compared the values obtained using the algorithm with the
ones known exactly for the case of the 4x4x4-ferromagnet.
The values of $\epsilon$ that I use are between $\epsilon=10^{-1}$
and $\epsilon=5\cdot 10^{-4}$. Certainly, $\epsilon$ should tend to zero
with increasing $t$ to obtain even more accurate values of the
entropy. However, the smaller $\epsilon$ is already in the
earlier stages of the algorithm, the longer it takes to
reach the asymptotic stage. If one takes $\epsilon$ too small at
the start of the algorithm, one risks to never leave, during the time of
the simulation, the regime where effectively one samples
according to performing a random walk in configuration space,
as the differences in the entropy values will stay too small
in the available simulation time. A good way to measure whether
one has reached the asymptotic regime of the algorithm yet or not
is to keep track of the sampled energies: if the histogram of the
energies, averaged over a long enough period of time, is flat,
asymptotics are reached. 

Some comments on the large-$\epsilon$ limit are also in place.
For $\epsilon$ of order $1$ or larger, the transition probability 
(\ref{tp}) of accepting a move becomes essentially $0$ or $1$, as the
differences in estimated entropy values for the different levels
become larger than $1$, hence the argument of the $\tanh$. This
has the following consequences. If, e.g., one is at time $t$, say, 
in a configuration of the level $\nu(t)$, and the values of the
estimated entropy values at the adjacent levels $\nu(t)\pm 1$
are $S(\nu(t)-1)<S(\nu(t))<S(\nu(t)+1)$, then moves are 
(essentially) accepted iff the level
of the considered update, $\nu_{\mbox{\tiny update}}$, is
$\nu_{\mbox{\tiny update}}=\nu(t)-1$. The algorithm can be 
viewed as putting a brick of height $\epsilon$ at every time 
step onto a wall which is building up during the
process. If the height of the wall at the 
adjacent place (level) whereto the move at time $t+1$ 
is considered is lower, the move is accepted with probability $\sim 1$,
if the wall is of equal height, the move is accepted with probability
$\frac{1}{2}$, and else it is (essentially always) rejected. 
It is easy to notice that on average the wall will be of equal
height for all levels if one substracts a running (increasing) baseline.
This leads to the observed fact that all energy levels occur
with equal probability, albeit the fact that the weight factors
in the algorithm are not proportional to the true entropy 
values (on average).

\subsection{The infinite-range ferromagnet}

I have considered systems which contained $N=2^n$ spins where
$n=2,3,\ldots,9$. I have compared runs where the ``tunneling
times'' where measured from the beginning with ones where the
tunneling times of the first $x\cdot{\cal N}^2$ ($x=1,2,\ldots,10$)
Monte Carlo steps (MCS; defined as usual as the time needed to
update all $N$ spins in the system) were not taken into 
account. The (expected) experience from
these runs is that the distribution of tunneling times, its
mean ($\tau_{\mbox{\tiny m}}$) and random mean square (rms) value
($\tau_{\mbox{\tiny rms}}$) remained essentially 
unaltered. However, to be on the safe side, I have not
counted the tunnelings observed during the first $10\cdot{\cal N}^2$ MCS
in the runs whose results are displayed in the following.
In all of the different runs, the histogram of the
energies, averaged at the same time than the (instantaneous) entropy
values, is flat, i.e., fluctuates around the mean number of
sampled configurations per energy to within at most a small 
fraction of a percent for small system sizes and up to 
at most $\pm 5\%$ for the largest systems considered.

The results shown in the figures have been obtained with a total
number of $10^6+10\cdot{\cal N}^2$ MCS. I have fixed $\epsilon$
to equal $10^{-3}$, $10^{-2}$, and $10^{-1}$, respectively. This implies
that the values $\tau_{\mbox{\tiny m}}$ and $\tau_{\mbox{\tiny rms}}$,
the mean and the rms value of the tunneling times, have been obtained
for $\epsilon=10^{-3}$ from 149338 values for $n=4$ to 659 values for
$n=9$, for $\epsilon=10^{-2}$ from 148726 values for $n=4$ to 1218 
values for $n=9$, for $\epsilon=10^{-1}$ from 146682 values for 
$n=4$ to 1826 values for $n=9$. To check for statistical 
reliability of the data, I have also performed slightly 
shorter and longer runs. The error bars that I have got from 
these runs, at fixed averaging times (of the instantaneous entropy),
are smaller than the size of the symbols in the figures. 
The distributions for the tunneling times, $\tau$,
themselves are, however, intrinsically very broad 
(their width is of the order of their mean)
with an accordingly long tail, possibly power-law-like. 
This can be seen in figure~1, where
the distribution, $\mbox{P}(\tau)$, of the ``tunneling times'', $\tau$, 
is shown for a run of $1002560$ MCS and $\epsilon=10^{-2}$
with $N=2^{32}$ spins. In total, 14941 tunnelings have been observed
during the last $10^6$ MCS, therefrom 45 events with a tunneling time
larger than $10\cdot\tau_{\mbox{\tiny m}}=10\cdot 1071$ MCS.
\begin{figure}
\makebox[0cm]{}
        \epsfxsize=14cm	 
        \epsfbox{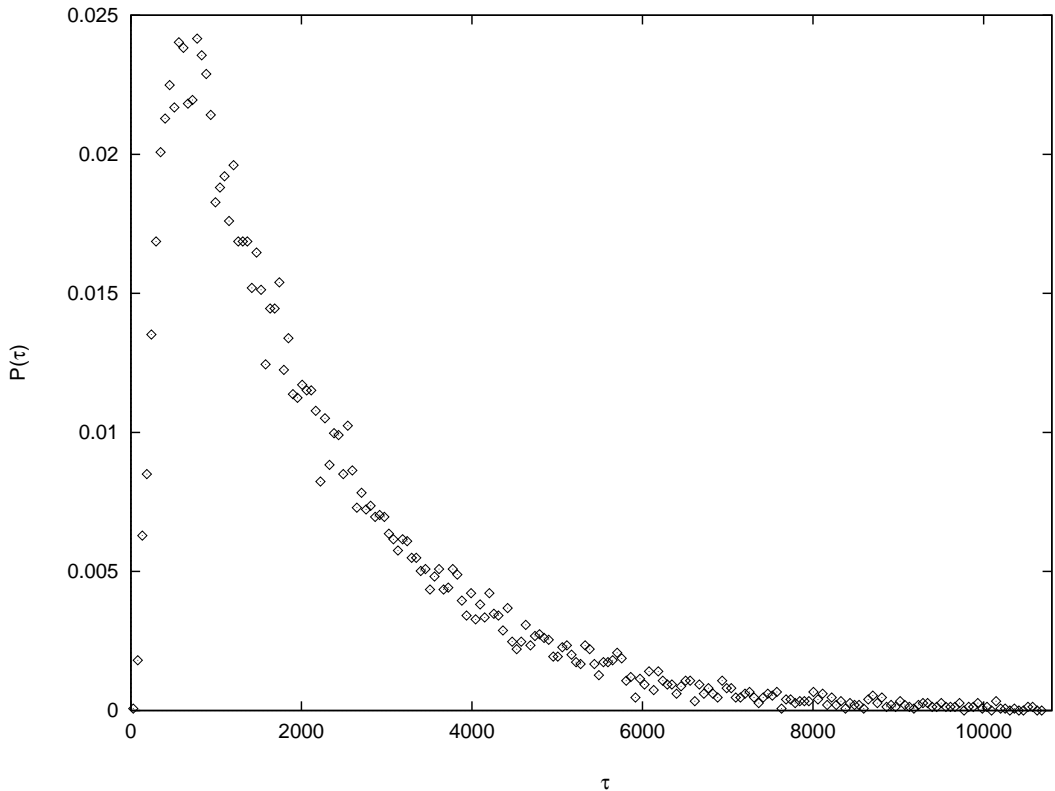}
\vspace{1cm}
\caption 
{Plot of the distribution, $\mbox{P}(\tau)$, of the ``tunneling times'', $\tau$,
for a run of $10^6$ MCS and $\epsilon=10^{-2}$ with $N=2^{32}$ spins.}
\end{figure}
In the runs displayed in the figures of this subsection, at most $432$ 
tunneling events with times above $5\cdot\tau_{\mbox{\tiny m}}$ occured
for $N=2$, this number rapidly decreasing with system size to at most
$2$ such events for $N=512$. The statistics for
the larger systems may, indeed, be insufficiant because of the long tail
of the distribution (the effect of a reduction of the measured 
$\tau_{\mbox{\tiny m}}$ can clearly be seen in figure~2 for the
runs with $\epsilon=10^{-2}$ and $256$ respectively $512$ spins). 
However, the resulting error should not alter the results significantly.

Figure~2 shows the increase of $\tau_{\mbox{\tiny m}}$ with the
number of the spins in the system on a log--log scale. 
\begin{figure}
\makebox[0cm]{}
        \epsfxsize=14cm	 
        \epsfbox{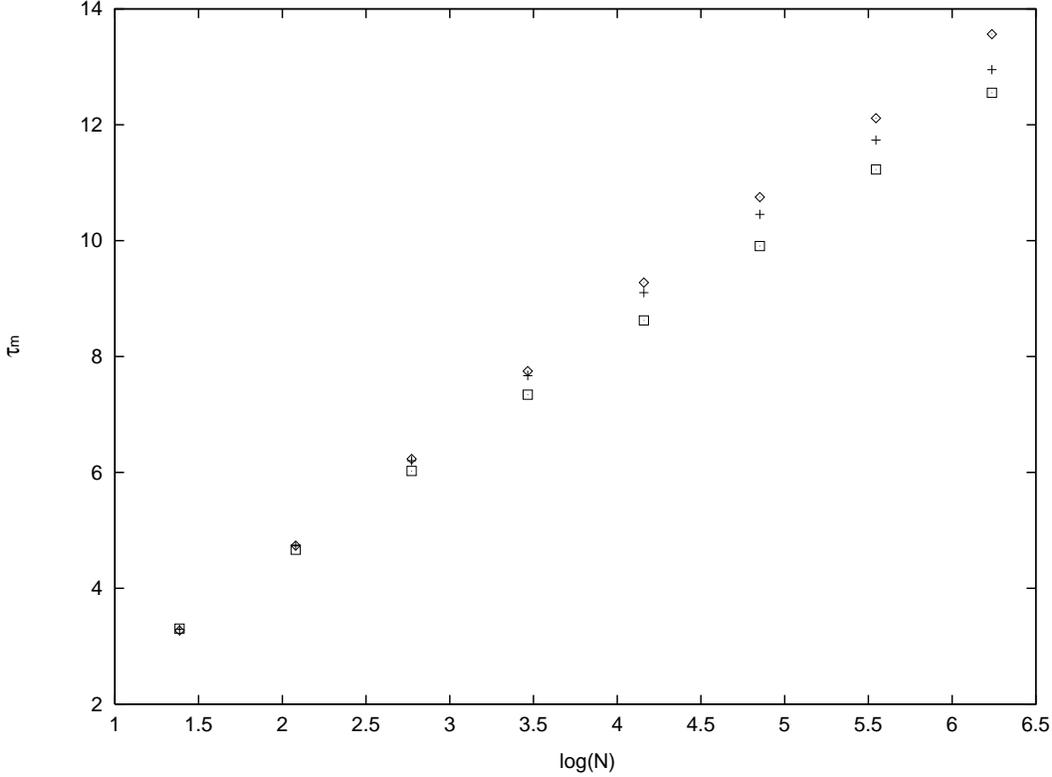}
\vspace{1cm}
\caption 
{Log-log-plot of the mean ``tunneling times'' 
$\tau_{\mbox \tiny m}$ versus the number of spins, $N=2^n$
($n=2,3,\ldots,9$), in the system for the infinite-range 
ferromagnet. The diamonds stem from simulations with $\epsilon=10^{-3}$,
the crosses from ones with $\epsilon=10^{-2}$, and the open squares
from ones with $\epsilon=10^{-1}$.}
\end{figure}
I have fitted straight lines (by eyesight) to the data points in figure~2, 
corresponding to the fits $\tau_{\mbox{\tiny m}}=c_{\mbox{\tiny m}} 
N^{\delta_{\mbox{\tiny m}}}$, for the different values of $\epsilon$. 
Because of higher statistical reliablitity of the results of the
smaller size systems these have been taken account more than the ones
of the larger size systems. The results for the fit parameters are
\begin{equation} 
\begin{array}{lcllcl}
\ln (c_{\mbox{\tiny m}}) \approx 0.36 \,\, , \quad  
\delta_{\mbox{\tiny m}} \approx 2.10 \,\, ,\qquad
\epsilon=10^{-3}\,\, ,\\
\ln (c_{\mbox{\tiny m}}) \approx 0.42\,\, ,  \quad \delta_{\mbox{\tiny m}} 
\approx 2.07 \,\, ,\qquad
\epsilon=10^{-2}\,\, ,\\
\ln (c_{\mbox{\tiny m}}) \approx 0.54\,\, , \quad \delta_{\mbox{\tiny m}} 
\approx 1.99\,\, ,\qquad \epsilon=10^{-1}\,\, .\\
\end{array}
\end{equation}
Remember that the exponent for a random walk in energy space 
is $\delta = 2$ from the general results on one-dimensional random
walks. 

In figure~3 the increase of $\tau_{\mbox{\tiny rms}}$ with the
number of the spins in the system is displayed on a log--log scale. 
\begin{figure}
\makebox[0cm]{}
        \epsfxsize=14cm	 
        \epsfbox{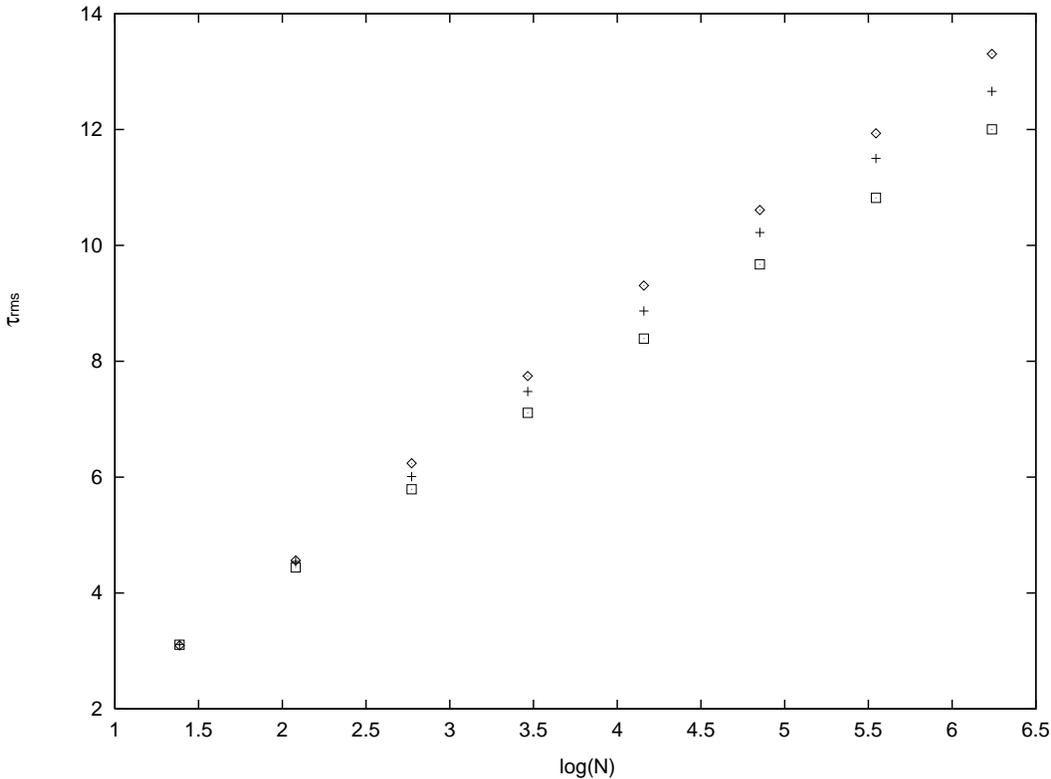}
\vspace{1cm}
\caption 
{Log-log-plot of the rms value of the ``tunneling times'' 
$\tau_{\mbox{\tiny rms}}$ versus the number of spins, $N=2^\nu$
($\nu=2,3,\ldots,9$), in the system for the infinite-range 
ferromagnet. The diamonds stem from simulations with $\epsilon=10^{-3}$,
the crosses from ones with $\epsilon=10^{-2}$, and the open squares
from ones with $\epsilon=10^{-1}$.}
\end{figure}
Here the straight lines corresponding to the fits
$\tau_{\mbox{\tiny rms}}=c_{\mbox{\tiny rms}} N^{\delta_{\mbox{\tiny
rms}}}$ give
\begin{equation} 
\begin{array}{lcllcl}
\ln (c_{\mbox{\tiny rms}}) \approx 0.15\,\, ,\quad  
\delta_{\mbox{\tiny rms}} \approx 2.12\,\, ,\qquad
\epsilon=10^{-3}\,\, ,\\
\ln (c_{\mbox{\tiny rms}}) \approx 0.23\,\, ,\quad  
\delta_{\mbox{\tiny rms}} \approx 2.08\,\, ,\qquad
\epsilon=10^{-2}\,\, ,\\
\ln (c_{\mbox{\tiny rms}}) \approx 0.36\,\, , \quad
\delta_{\mbox{\tiny rms}} \approx 1.98\,\, ,\qquad
\epsilon=10^{-1}\,\, .\\
\end{array}
\end{equation}
We see from the figures that the width of the distribution 
is of the order of its mean, which can be noted easily already for
$N=1$, where one knows the (power-law-like) distribution of the tunneling
times and its properties analytically. 
 
In all of the above simulations, I have compared the obtained values
for the entropy with the true ones, from equation (\ref{defomega}). Depending 
on the value of $\epsilon$, on the number of times that I averaged
over the (estimated instantaneous) entropy values and on the 
total time of the simulation, I got more and less accurate results.
In any case, for the runs of figures~2 and~3, the error was typically
of the order of $\epsilon$. In figure~4, a comparison of the entropy
values, obtained for a system containing $128$ spins, using the algorithm
with $\epsilon=10^{-2}$ and running it for a total of 50960 MCS, with 
the exact values is shown.
\begin{figure}
\makebox[0cm]{}
        \epsfxsize=14cm	 
        \epsfbox{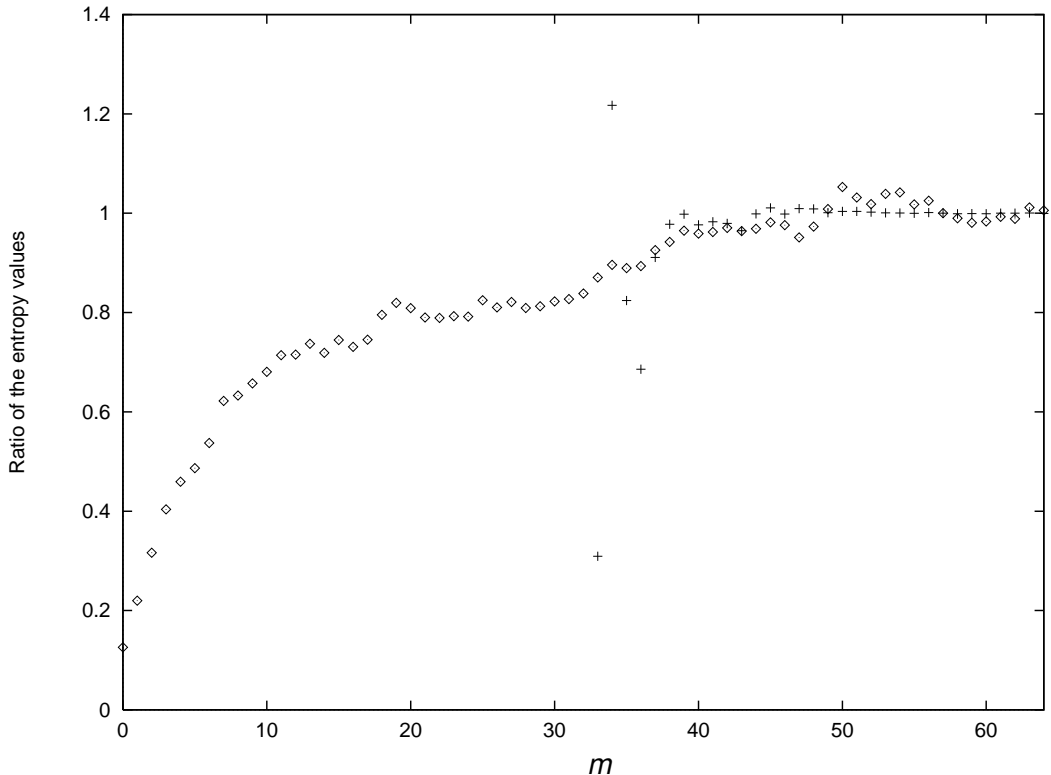}
\vspace{1cm}
\caption 
{Comparison of the entropy values, $S(m)$, obtained through the FEMC algorithm
and $\epsilon=10^{-2}$ (crosses) with the true entropy values, $S_0(m)$, and
comparison of the ones obtained through a random walk in 
configuration space (diamonds) with the true entropy values, 
for a system with 256 spins. The FEMC algorithm has been run in total for 
50960 MCS, the one of the random sampling for 1040960 MCS.}
\end{figure}
As can be seen from figure~4, the matching of the values (of the number
of configurations, not of the entropy~!) is quite
impressive, taking into account the fact that there are $2^{128}$
configurations. I have also compared it in the figure to the values 
obtained by a run with a larger number of MCS, namely 1040960 MCS,
but performing a local random walk in configuration space. In this 
last run, I have counted the configurations during the run and normalised
the sum of all the hits to $2^{128}$. It is easily
seen from the figure that, in contrast to the FEMC algorithm (with
much fewer MCS~!), ergodicity has been lost, configurations with 
energy $E(m)$ for $m<33$ have not even been sampled~!

In conclusion, the FEMC algorithm provides for ergodicity of the
system in the used computer time, and satisfactory estimates 
of the entropy for $\epsilon$ in the range between $10^{-1}$ 
and $10^{-3}$. If one is rather interested in smaller tunneling 
times, one should however run the algorithm with larger $\epsilon$.

\subsection{The two- and three-dimensional ferromagnets}

In the case of the two-dimensional ferromagnet I have performed
runs on systems of linear size $L=2^n$, where $n=1,2,\ldots,6$,
and for the three-dimensional ferromagnet on systems of 
linear size $L=2^n$, where $n=1,2,3,4$. 
Again, I have compared runs where the tunneling
times where measured from the beginning with ones where the
tunneling times of the first $x\cdot{\cal N}^2$ ($x=1,2,\ldots,10$)
MCS were not taken into account and observed 
that the distribution of the tunneling times, its
mean and rms value remained essentially 
unaltered. However, to be again on the safe side, I have not
counted the tunnelings observed during the first $10\cdot{\cal N}^2$ MCS
in the runs whose results are displayed in the following.
In all of the different runs, the histogram of the
energies, averaged at the same time than the (instantaneous) entropy
values, is flat to within the same percentages as in the infinite-range
ferromagnet.

The results shown in figure~5 (for the two-dimensional ferromagnet)
have been obtained with a total number of $10^6+10\cdot{\cal N}^2$ MCS. 
As these runs were done mainly to test the tunneling ability of the
algorithm, I have fixed $\epsilon$ to equal $10^{-1}$. This implies
that the values $\tau_{\mbox{\tiny m}}$ and $\tau_{\mbox{\tiny rms}}$,
the mean and the rms value of the tunneling times, have been obtained
from 106825 values for $L=2$ to 14 values for
$L=64$. The same statistical errors as for the infinite-range 
ferromagnet play a r\^ole here. 
The distributions for the tunneling times, $\tau$,
themselves are, again, intrinsically very broad 
(their width is of the order of their mean~!)
with an accordingly long tail. 
In the runs displayed in the figure $393$ tunneling events occured
with times larger than $5\cdot\tau_{\mbox{\tiny m}}$ for $L=2$ down to
$1$ for $L=64$. The statistics for
the larger systems may again be insufficiant because of the long tail
of the distribution.

Figure~5 shows the increase of $\tau_{\mbox{\tiny m}}$ and 
$\tau_{\mbox{\tiny rms}}$ with the
volume of the system on a log--log scale. 
\begin{figure}
\makebox[0cm]{}
        \epsfxsize=14cm	 
        \epsfbox{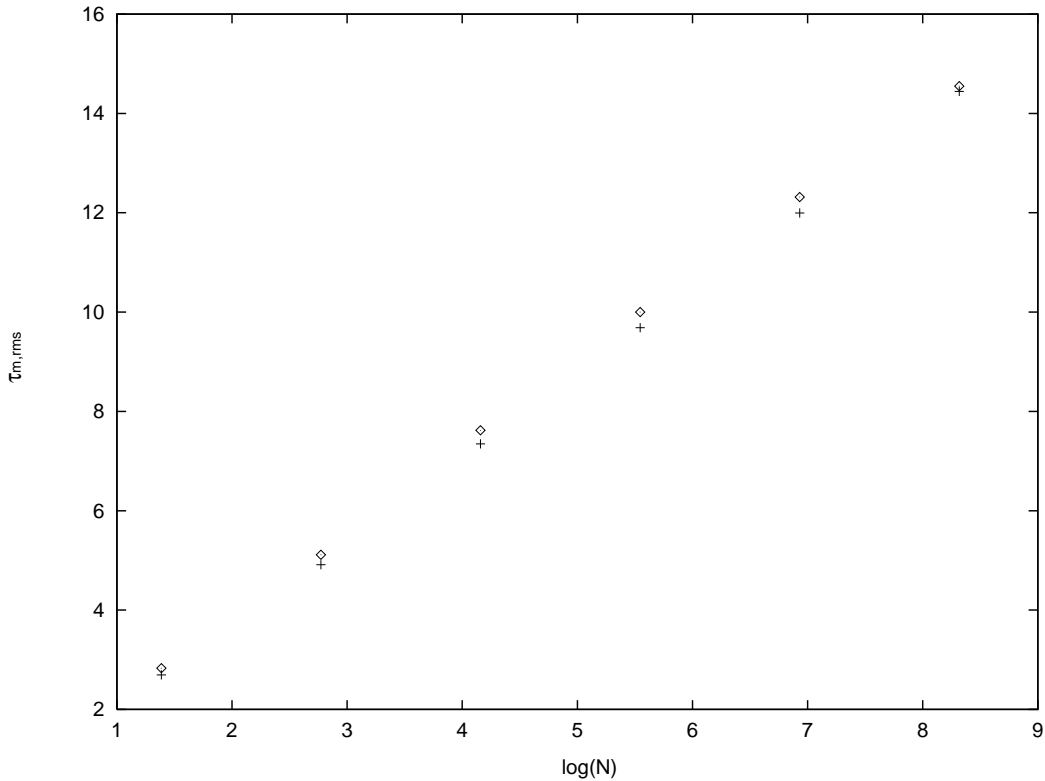}
\vspace{1cm}
\caption 
{Log-log-plot of the mean ``tunneling times'' 
$\tau_{\mbox \tiny m}$ (crosses) and of their rms value 
$\tau_{\mbox{\tiny rms}}$ (diamonds) 
versus the volume size of the system, $V=(2^n)^2$
($n=1,2,\ldots,6$), for the two-dimensional ferromagnet.}
\end{figure}
I have fitted straight lines (by eyesight) to the data points in figure~5, 
corresponding to the fits $\tau_{\mbox{\tiny m}}=c_{\mbox{\tiny m}} 
N^{\delta_{\mbox{\tiny m}}}$ . The results for the fit parameters are
\begin{equation} 
\begin{array}{lcllcl}
\ln (c_{\mbox{\tiny m}}) &\approx& 0.36 \,\, , \quad  
&\delta_{\mbox{\tiny m}} &\approx& 2.10 \,\, ,\\
\ln (c_{\mbox{\tiny rms}}) &\approx& 0.42\,\, ,  \quad 
&\delta_{\mbox{\tiny m}} &\approx& 2.07 \,\, .
\end{array}
\end{equation}
We notice again from the figure that the width of the distribution 
is of the order of its mean.

The results shown in figure~6 (for the three-dimensional ferromagnet)
have also been obtained with a total
number of $10^6+10\cdot{\cal N}^2$ MCS and $\epsilon=10^{-1}$. 
This implies that the values $\tau_{\mbox{\tiny m}}$ and 
$\tau_{\mbox{\tiny rms}}$ have been obtained from 443 values 
for $L=2$ down to 6 values for $L=16$.  
In the runs displayed in the figure, 3 tunneling events occured
with times larger than $5\cdot\tau_{\mbox{\tiny m}}$ for $L=2$ down to
$0$ for $L=16$. The statistics for
the larger systems may again be insufficiant because of the long tail
of the distribution. Figure~6 shows the increase of $\tau_{\mbox{\tiny m}}$ and 
$\tau_{\mbox{\tiny rms}}$ with the volume of the system on a log--log scale. 
\begin{figure}
\makebox[0cm]{}
        \epsfxsize=14cm	 
        \epsfbox{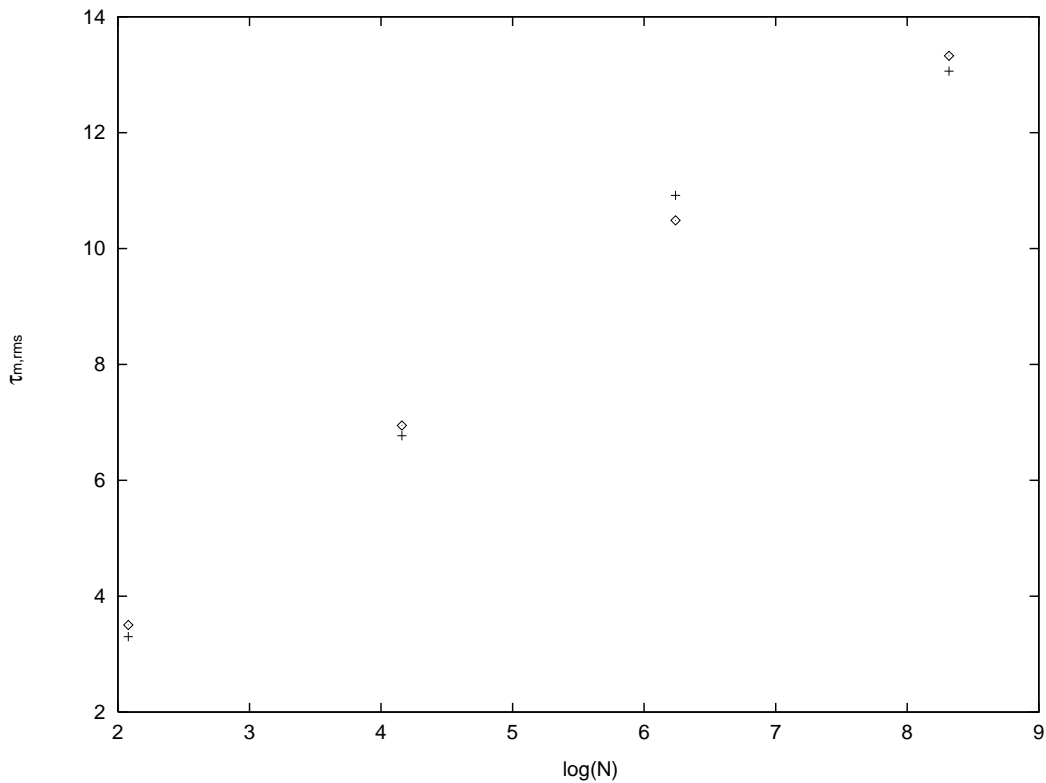}
\vspace{1cm}
\caption 
{Log-log-plot of the mean ``tunneling times'' 
$\tau_{\mbox \tiny m}$ (crosses) and of their rms value 
$\tau_{\mbox{\tiny rms}}$ (diamonds) 
versus the volume size of the system, $V=(2^n)^3$
($n=1,2,\ldots,4$), for the three-dimensional ferromagnet.}
\end{figure}

Finally, I have also determined approximate values of the
entropy for the 4x4x4-ferromagnet in order to compare them
to the ones known exactly \cite{P82}. The resulting comparison
can be seen in figure~7. The FEMC simulation of a total of $120000$ MCS
(of which the first 100000 MCS were discarded from the measurements;
during the last 20000 MCS the entropy was averaged over the
instanteneous values of every 100th MCS)
has been done with fixed $\epsilon=5\cdot 10^{-4}$ .
\begin{figure}
\makebox[0cm]{}
        \epsfxsize=14cm	 
        \epsfbox{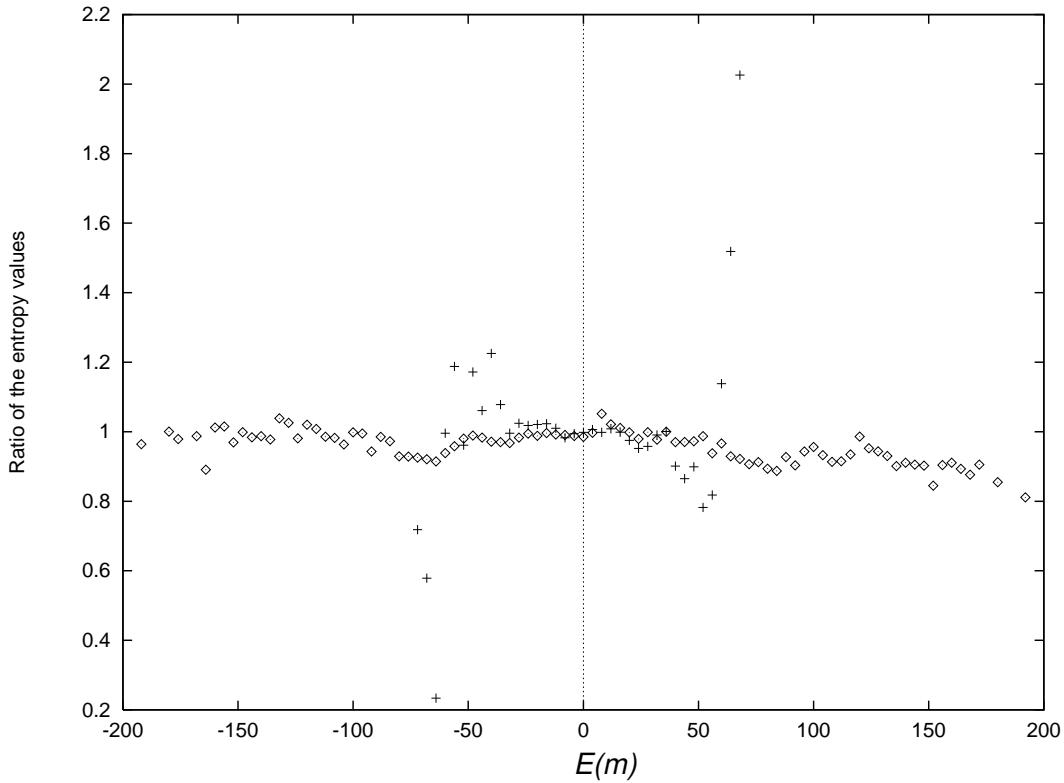}
\vspace{1cm}
\caption 
{Comparison of the ratio of the number of configurations from 
FEMC simulation/exact values (diamonds) for the 4x4x4-dimensional ferromagnet 
from a run with a total of $120000$ MCS and of the ratio of the number
of configurations from a simulation through random walk in configuration
space/exact values (crosses) for $10^6$ MCS.}
\end{figure}
As can be seen from the figure, the matching of the values (of the
number of configurations, not of the entropy~!) is again 
quite impressive, taking into account the fact that there are $2^{64}$
configurations. I have also compared it in the figure to the values 
obtained by a run performing a local random walk in configuration space,
of even more MCS, namely $10^6$ MCS. In this 
latter run, I have counted the configurations sampled during the run and normalised
the sum of all the hits to $2^{64}$. It is easily
seen from the figure that, in contrast to the FEMC algorithm, 
ergodicity has again been lost, configurations with 
energies $|E|>88$ have not even been sampled~!

In conclusion, the FEMC algorithm provides also for the two- and
three-dimensional ferromagnets for ergodicity of the
system in the used computer time, 
and satisfactory estimates of the entropy for $\epsilon$
in the range between $10^{-1}$ and $5\cdot 10^{-4}$. If one is rather
interested in smaller tunneling times, one should however again run
the algorithm with larger $\epsilon$.

\section{Conclusion}

Recently Monte Carlo sampling with respect to unconventional ensembles
has received some attention [9--34]. Multicanonical and related sampling has 
allowed considerable gains in situations with ``supercritical'' slowing 
down, such as first order phase transitions \cite{our1,our1a,our4,Janke}
and systems with conflicting constraints, such as spin glasses
\cite{our2a,temp,BHC,Ker1} or proteins \cite{HO,HS}. They seem, however,
still be haunted by some difficulties, in particular for the latter
systems \cite{B95}.

For first order phase transitions of non-random systems 
the problem of the a--priori unknown
entropy is rather elegantly overcome by means of finite size
scaling methods \cite{our1,our1a,our4,Julich,BNB,Janke}. A sufficiently
accurate estimate is obtained by extrapolation from the already simulated
smaller lattices. The smallest lattices allow still for efficient
canonical simulations. For the three-dimensional Ising ferromagnet 
it is clear that the finite size scaling methods employed
in \cite{our1,our1a,our4} provide reliable estimates of the multicanonical
parameters. On the
other hand, this approach breaks down \cite{our2a} for the important
class of disordered systems. For instance for spin glasses one has to perform the
additional average over quenched random variables (the coupling
constants). Different choices of these random variables define
different realisations of the (finite-size) system which imply different
entropy functions for the different realisations.
Then algorithms like the one of this paper
become crucial, but the Ising ferromagnet is still
a suitable testing ground to set quantitative performance
scales. 

In this paper, I have introduced a new algorithm in the above spirit.
I have analytically proven that the entropy of a system can be 
obtained from a {\em Free Energy Monte Carlo}\/ algorithm at 
infinite temperature ($\beta=0$ in equation (\ref{FEMC})) in the 
infinite-time limit and applied the algorithm to the infinite-range
as well as to the two- and three-dimensional ferromagnets. 
I have simulated these systems to obtain the entropy function 
and to investigate the ``tunneling times'' of getting from one
energy minimum to another one. The results are very encouraging: 
\begin{description}

\item{(i)} The scaling of the ``tunneling times'' with the system size 
are almost the ones of a random walk in energy space. 

\item{(ii)} The entropy function of the considered systems could be obtained
roughly within errors of order $\epsilon$ in the used computer time. 

\item{(iii)} More importantly, ergodicity could be retained in all the
considered cases in the used computer time ($\sim 10^6$ MCS at most).

\end{description}
In particular, the last fact, the retaining of the ergodicity, should allow
for a calculation of physical quantities, such as correlation
functions, through equation (\ref{erg}) near zero temperature 
where conventional MC simulations fail. It is also interesting
to use the full FEMC algorithm (with nonvanishing $\beta$). More
results with the latter shall be published elsewhere \cite{t97}.
It is particularly interesting to determine the best choice of the
parameters $\epsilon$ and $\beta$ and of their ratio to overcome
energy barriers in the least amount of time and to explore the energy
space for minima or more generally extrema. Also, more 
analytical details and properties of the algorithm can be obtained 
for the case of the infinite-range ferromagnet \cite{t97}. It shall
be interesting to apply the algorithm in the cases where hitherto
ergodicity problems have not allowed to obtain results.

One last application of the algorithm should be mentioned here.
Constructing a general model of on-line learning 
is an important challenge in the theory of learning and its application. 
A plausible definition of the goal of supervised learning from examples is
to find a weight vector $\bf{w}$ that minimises the generalisation error, 
$\epsilon_g (\bf{w})$.
A general model of batch learning, 
in which the learner has free access to a fixed 
set of examples, is based on minimisation of 
the total training error. Indeed, as the size of the training set, $P$,
grows this procedure converges uniformly to the minimum of the 
generalisation error.
In systems with continuously varying weights,
the rate of convergence to this limit follows generically
a power law \cite{sst-smle-92}.
A similar general model does not yet exist for on-line learning,
where, at each time step, the learner receives a single new
example and is unable to store previous examples in
memory. 
The conventional on-line algorithm is based on the gradient of
the instantaneous error \cite{Hybook,hk-lpnn-91}.
For a sufficiently small learning rate, it converges
to a local minimum of  $\epsilon_g (\bf{w})$ but not necessarily to 
the global one. More importantly, it is not applicable to learning 
of boolean functions or of other discrete valued functions which are extremely 
useful for decision and classification tasks. 
Recently, an {\em on-line Gibbs algorithm}\/ has been proposed \cite{SK96}
as the first on-line algorithm that guarantees convergence 
to the minimal generalisation error for 
non-smooth systems, in particular for 
systems with discrete valued outputs or threshold hidden units. 
The price that is paid is an increased complexity of the computation at
each presentation of examples.
In particular, for systems in which the generalisation error has local minima, 
the on-line Gibbs algorithm may require a slow 
annealing schedule of the temperature variable, used in the algorithm,
 which might yield a slow global convergence rate. 
In addition, the algorithm relies on the 
possibility of updating the weights by small increments. Consequently,
it is inapplicable to systems with {\em discrete valued weights}\/. 
In the case of the FEMC algorithm, gradient descent and escaping from 
minima can be combined naturally (see equation~(\ref{FEMC})) 
so that the FEMC may offer a valuable alternative to optimise the
learning procedure. Serious questions concerning 
on-line learning are still open, most importantly the one is on the 
{\em global convergence rate}\/, which is yet unknown in general.

\end{document}